# Accelerating Test Automation through a Domain Specific Language

Anurag Dwarakanath[1], Dipin Era[1], Aditya Priyadarshi[2], Neville Dubash[1] & Sanjay Podder[1]

[1] Accenture Technology Labs
Bangalore, India

[2] College of Computer and Information Science
Northeastern University
Boston, USA

*Abstract*—Test automation involves the automatic execution of test scripts instead of being manually run. This significantly reduces the amount of manual effort needed and thus is of great interest to the software testing industry. There are two key problems in the existing tools & methods for test automation - a) Creating an automation test script is essentially a code development task, which most testers are not trained on; and b) the automation test script is seldom readable, making the task of maintenance an effort intensive process. We present the Accelerating Test Automation Platform (ATAP) which is aimed at making test automation accessible to non-programmers. ATAP allows the creation of an automation test script through a domain specific language based on English. The English-like test scripts are automatically converted to machine executable code using Selenium WebDriver. ATAP's English-like test script makes it easy for non-programmers to author. The functional flow of an ATAP script is easy to understand as well thus making maintenance simpler (you can understand the flow of the test script when you revisit it many months later). ATAP has been built around the Eclipse ecosystem and has been used in a real-life testing project. We present the details of the implementation of ATAP and the results from its usage in practice.

## I. INTRODUCTION

Software testing is considered as one of the most expensive phases of the entire software development cycle. Estimates have put the amount of effort needed for testing to be between 30% & 90% [2] of the overall software development effort. Automation in testing, thus, is seen as a way to reduce the amount of manual effort needed. A prominent approach for automation is the execution of test scripts by a machine instead of being manually run. There have been various tools developed for the automatic execution of test scripts including Selenium WebDriver [14], HPE Unified Functional Testing (UFT) [5] and IBM Rational Functional Tester (RFT) [6]. However, the adoption of such tools in practice is miniscule and projects have reported little or no benefits of automation [20][7][12].

The poor adoption is due to two fundamental properties of the current techniques of test automation - a) The creation of automation scripts is essentially a code development task and is typically difficult for non-programmers (such as testers) [12] [10]; b) Even if the initial automation scripts have been made by developers, the maintenance of such automation scripts is costly as the scripts are difficult to read and comprehend by non-programmers.

Our aim is to build a tool that can make test automation accessible to non-programmers. We present the Accelerating Test Automation Platform (ATAP) which allows a tester to author automation test scripts in an English-like language. Figure 1 shows a comparison of our method with that of Selenium WebDriver. The scenario requires the text of "New York" to be entered into the textbox of the Expedia webpage. In the case of WebDriver, the automation script is coded in Java. However, in the case of ATAP, the same functionality is expressed in a much more readable way where significant verbiage of Java is avoided.

ATAP has been built around the Eclipse ecosystem. We have developed a domain specific language (DSL) which forms the grammar of ATAP (i.e. keywords used to author automation scripts). The DSL has been built using the Xtext [4] and Xbase [3] frameworks and follows the Gherkin [19] syntax. Statements written through this grammar are then converted into Java code using Selenium WebDriver. The code generation is written in the Xtend [1] programming language. The generated Java code then executes the test step.

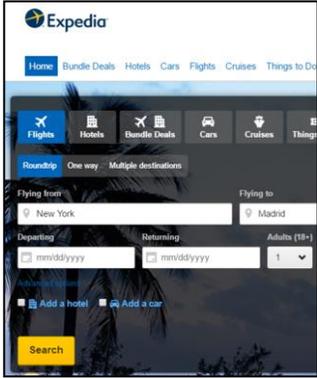

Figure 1. Authoring test scripts in an English-like language.

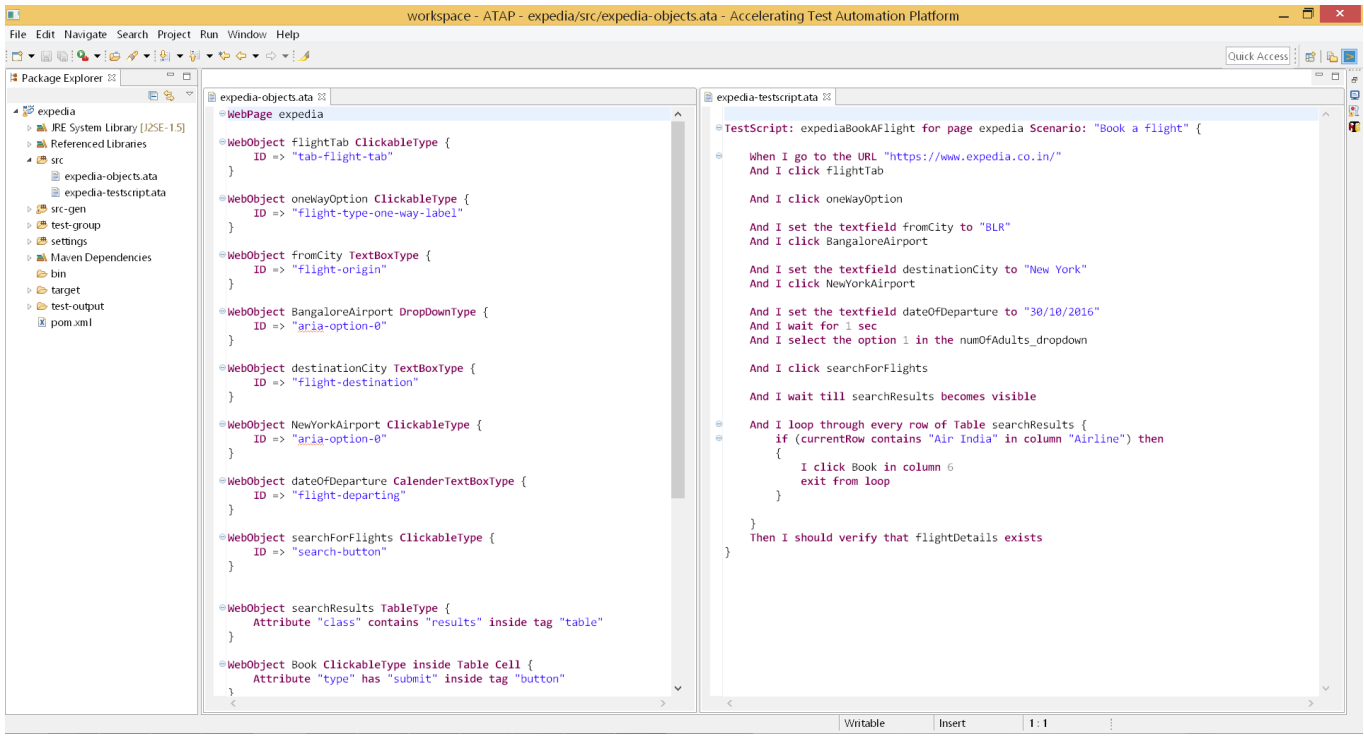

**Figure 2. View of the ATAP tool. 2 files need to be authored by a tester – a) the Objects file (left pane) and b) The test steps (right pane).**

Figure 2 shows the view of ATAP to author an automation script. The script searches for flight options between two cities and clicks on 'book' for a particular airline. The left pane of the view displays objects file (called as `expedia-objects.ata`) and the right pane shows the test steps (called `expedia-testscript.ata`). Notice that the test steps, in the right pane, is self-explanatory in its functional flow. This is the fundamental value of using ATAP, where the English-like scripts allows a tester to easily create and readily comprehend the flow of the test script.

ATAP has been used in an industrial software testing project. A set of test scenarios, ranging in complexity, was automated using ATAP. ATAP was used by three groups of people – a) the automation engineers of the project (who can be considered as experts in Selenium, but new to ATAP); b) the manual testers of the project (who can be considered new to automation and new to ATAP) and c) the developers of ATAP (who can be considered as ATAP tool experts). We compared the coverage of ATAP (whether it can cater to all the scenarios) and measured the amount of time it took to create the automation scripts.

We found that ATAP could cover all of the scenarios needed by the project. The automation engineers and the ATAP tool developers could use ATAP to completely automate the test scenarios. On average, automating through ATAP showed a 25% effort savings when compared to Selenium WebDriver. The manual testers, on average, could automate 71% of the test scenario.

We also observed and received various qualitative feedback from the project team. We have presented these learnings in Section V.

This paper is structured as follows. We present the related work in the space of test automation in Section II. The architecture and implementation details of ATAP are presented in Section III. We present the results from the usage of ATAP in an actual project in Section IV. The lessons learnt are presented in Section V and we conclude in Section VI.

## II. RELATED WORK

The tools & techniques in the automated execution of test scripts for web based systems can be broadly classified into 4 types: a) Programming based automation; b) Record & Replay; c) Image Recognition based automation and d) Natural Language based automation.

### A. *Programming based automation*

Programming based approach to test automation requires the creation of code through programming languages which can interact with the web elements in the web browser. Prominent tools using this approach include Selenium WebDriver [14], HPE Unified Functional Testing (UFT) [5] and IBM Rational Functional Tester (RFT) [6].

Figure 1 shows the code using WebDriver that needs to be written to interact with a textbox. The typical challenge with this method of automation is that the creation of automation test scripts becomes a code development task [12] and is not in the realm of software testers [10]. Typical software testers, who bring knowledge of the application and the knowledge of testing, are averse to such programming – i.e. while testers can present the test conditions that are needed, the actual creation of the test automation would need to be done by programmers.

To address the differing focus of testers and programmers, a tool & method called Cucumber [19] was developed. In Cucumber, the test conditions (i.e. the high level description of the test script – something similar to the 'scenario' shown in Figure 1) are created by the tester. The actual automation script using code is then done by the programmer [10]. Cucumber provides a guideline on the language to be used in the test conditions. The guideline follows the Gherkin syntax [19] which uses the keywords of 'Given', 'When' and 'Then'. In ATAP, we follow a syntax similar to Gherkin, where the English-like test scripts are written through the 'Given/When/Then' keywords.

Our approach with ATAP is to make automation accessible to non-programmers. In ATAP, a tester can author an automation test script without using any programming language. The test script is authored using an English-like syntax and the tool automatically converts the statements into Java code.

### B. Record & Replay

The record & replay method allows a tester to create an automation test script by performing the steps manually for the first time. In this record phase, the tool picks up the implementation specific details of the application as the tester performs the actions manually. The recorded test script is then automatically executed by the machine in the replay phase. Selenium IDE [13] and ATA-QV [20] are some of the tools which follow record & replay paradigm.

Record & replay makes it extremely easy to create an automation test script where a tester without needing the knowledge of programming can create an automation test script. However, the method leads to severe problems in maintenance [17]. The recorded test script is not modularized and is akin to having a complete program in a single class file. A single update to a web object identifier may require the update of multiple recorded test scripts. Further, to update a particular test step, the tester may require the re-recording of the entire test scenario from scratch.

It has been seen that the usage of record & replay results in more manual effort eventually (due to maintenance) than the programming based approach [8] [12].

### C. Image Recognition based approach

Image recognition based automation tools allows a tester to write a test step with the web element represented as an image. The tester first captures the image of the web element as displayed in the GUI and uses it in the test script. To execute such a test step, the tool locates that portion of the GUI which is similar to the image provided by the tester. This identification is done based on the intensity of the pixels (i.e. the code representing the pixels). Sikuli [15] is a popular tool using this paradigm.

The image recognition based method identifies web elements through a pixel to pixel comparison between two images. Such techniques have been found to take more time to create [9] than programming based automation since images to interact with every action need to be taken (for example, selecting a value from a dropdown needs two images to be taken – an image to click on the dropdown and a second image to choose the value). Image recognition based techniques have been shown to be extremely brittle needing significant maintenance [9] as small changes to the application (such as a change in the browser resolution or a change in the background color of a page) can break the automation script.

### D. Natural language based approach

The natural language based automation approach allows the usage of natural language test scripts to author test automation. A prominent tool employing this approach is ATA [17].

In ATA, the test scripts are written in natural language and are parsed into tuples of action & web objects. The web objects (which are described in a natural language) are then located in the web page by finding that element in the html which is close to the natural language description. Here 'closeness' is measured in terms of lines of code in the html of the webpage. For example, when the natural language script is written as 'Flying from textbox' (refer to Figure 1), ATA locates a web element in the html that is close to the text of 'Flying from'. This web element is then interacted upon based on the action.

We believe a key problem with the method of ATA is the usage of natural language test scripts. Natural language is known to be inherently ambiguous. Often, the intent of the script is difficult to decipher and thus the method fails frequently in practice. The study of the ATA method inspired us to attempt a different direction – the usage of a DSL for test automation. The DSL allows for an unambiguous interpretation of the test script and thus is far more applicable in practice.

### III. ACCELERATING TEST AUTOMATION PLATFORM

The Accelerating Test Automation Platform (ATAP) allows a tester to author automation test scripts using a domain specific language (DSL). The DSL follows an English-like syntax making it easy to author and maintain the automation scripts. ATAP internally converts the statements written in the DSL into Java code using Selenium WebDriver APIs. The system architecture is shown in Figure 3.

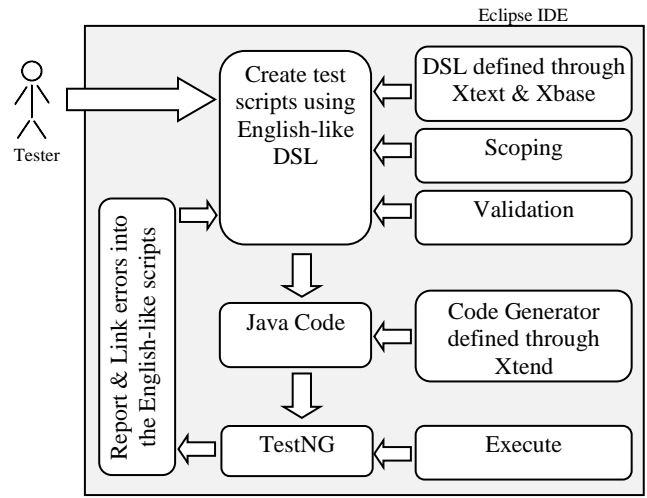

**Figure 3. Architecture of ATAP.**

## A. The Domain Specific Language

A fundamental property of ATAP is the ability to author automation test scripts in an English-like syntax. We have used the Xtext [4] and the Xbase [3] framework to build a DSL using which the automation commands can be authored.

A test automation script in ATAP consists of two files – one file where the objects are declared and the other where the test steps are declared. Figure 4 shows a web object by name 'fromCity' declared using our DSL. Keywords in the DSL are shown in **magenta** color. In Figure 4, the 'fromCity' web object is of a textbox type and is located in the webpage using the html attribute of ID.

```
WebObject fromCity TextBoxType {
    ID => "flight-origin"
}
```

**Figure 4. An Object declaration through the DSL.**

Figure 5 shows a test step authored using our DSL. The command depicted is the entry of the text "BLR" into the textbox defined in Figure 4. Notice the simplicity and descriptive nature of the test step.

```
When I set the textfield fromCity to "BLR"
```

**Figure 5. A test step declaration through the DSL.**

Different commands supported in our DSL for the object and test step declaration can be seen in Figure 2.

The use of Xtext and Xbase brings the feature of 'proposals'. Here, at every context, the system can provide the list of possible keywords that can be entered. Having such proposals, makes it easy for a tester to start authoring the test steps without having to memorize the keywords of our DSL. Figure 6 shows the proposals at the time of authoring an object. The figure shows the different object types supported in the DSL.

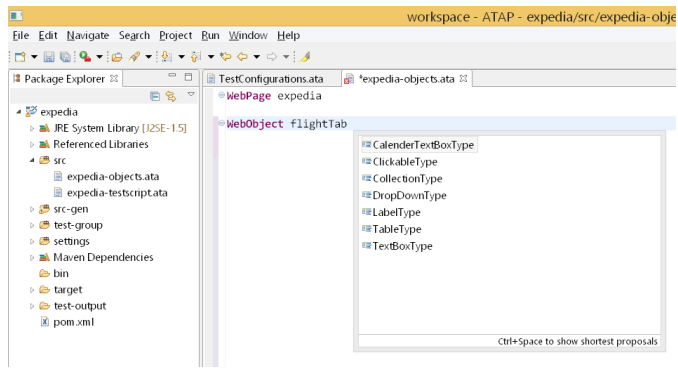

**Figure 6. Proposals when declaring a Web Object.**

Figure 7 shows the proposals at the time of creating a test step. The figure shows the various commands supported in the DSL. Through the use of the proposals, the tester can choose the desired option from the list as opposed to memorizing the syntax of the DSL.

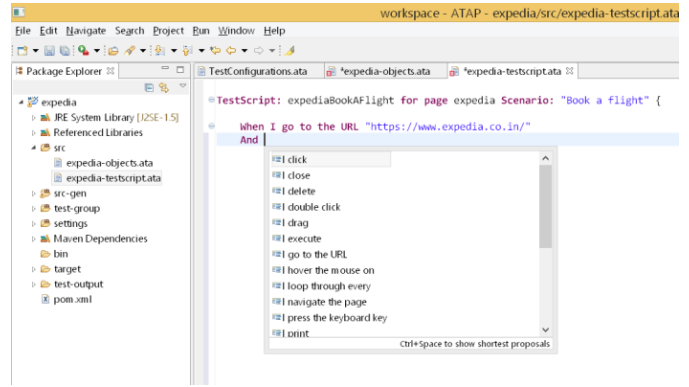

**Figure 7. Proposals when declaring a test step.**

As can be observed in Figure 2 and Figure 7, we have created the DSL to follow the Gherkin syntax [19] of 'Given-When-Then'. The keyword 'Given' denotes the set of pre-conditions that should be executed before the current test script can be executed. The keyword 'When' denotes the action that should be performed and the keyword 'Then' denotes the set of validations.

We have developed the DSL keeping two considerations in mind – a) The DSL should be as close as possible to a popular non-technical language such as English; and b) the DSL should be limited to the most used commands.

Aspect a) makes it easy for non-programmers to create an automation test script, where little or no learning is needed. It also allows one to quickly understand the flow of a test script so that maintenance is simpler (i.e. when one re-visits the test script after a few months of its creation, the intent of the script can be quickly deciphered).

Aspect b) allows a tester to choose the command from a list through proposals. There are numerous commands available in Selenium WebDriver and if we build a DSL mapping every command in WebDriver with an English-like statement, the number of options that are shown to the user would be extremely large and potentially confusing. We thus seek to build the DSL with a limited set of commands that are commonly used. We went through multiple test scripts and spoke to various testers to capture this common set of test actions. To achieve coverage (i.e. to ensure that any command supported in WebDriver should be possible in ATAP), we will build methods to invoke any Selenium WebDriver APIs (and Java statements) directly from the DSL statements (and explained subsequently).

## B. Invoking Java Code from the DSL

At the time of the design of the DSL, we chose to represent only the most common Selenium commands in an English-like format. This, then, requires a mechanism where non-represented commands can be invoked through the DSL, to ensure there is complete coverage.

We have included a mechanism to invoke any Java command directly from the DSL and shown in Figure 8.

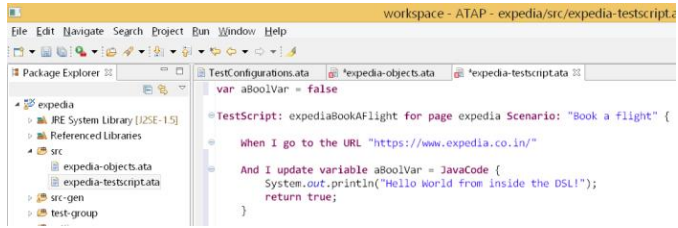

**Figure 8. Invoking Java Code through the DSL.**

The invocation of the Java code is allowed through the keyword '**JavaCode**' in our DSL. Internally, we use the XBloc expression syntax of the XBase framework. As can be seen in the figure, the Java code is expected to return a value of primitive datatype (Boolean, integer, or string) and this gets assigned to a variable declared in the DSL.

All variables and objects declared in the DSL (outside of the JavaCode scope) is also visible inside the scope for the declaration of the Java code.

*C. Scoping*

Through the use of scoping, ATAP restricts the options that are listed in the proposals. For example, when a tester chooses a command of '**I click**' (as shown in Figure 7), the list of web objects that are shown in the proposals are restricted to only those that are compatible for the action of 'click' – i.e. web objects that are defined as a 'TextboxType' are not shown when the 'click' action is selected. This allows the tester to create valid test steps at the time of authoring. In contrast, Selenium WebDriver allows such incompatibilities at the time of creating a test script and will generate a run-time exception upon execution.

Scoping has been used extensively in ATAP. We restrict the possibilities displayed during the creation of test steps, assignment of variables to values (i.e. type checking is done), etc.

*D. Validation*

Through the use of validation, ATAP is able to spot errors in the automated test script at the time of creation. For example, when a tester has used the 'Classname' selector as a method to identify a web object and supplies multiple words as the class name, ATAP raises an error. This is because the Selenium WebDriver API treats a class name with multiple words as a compound class. This will typically lead to a run-time exception in Selenium WebDriver. In ATAP, this error is caught at the time of creation. ATAP also gives a quick fix where the correct way to identify a web object with a compound class name is automatically created. This is shown in Figure 9.

Similarly, other such best practices have been programmed into the validation feature of ATAP. This includes spotting cases where the same web object has been declared multiple times, when incorrect path locations are used, etc.

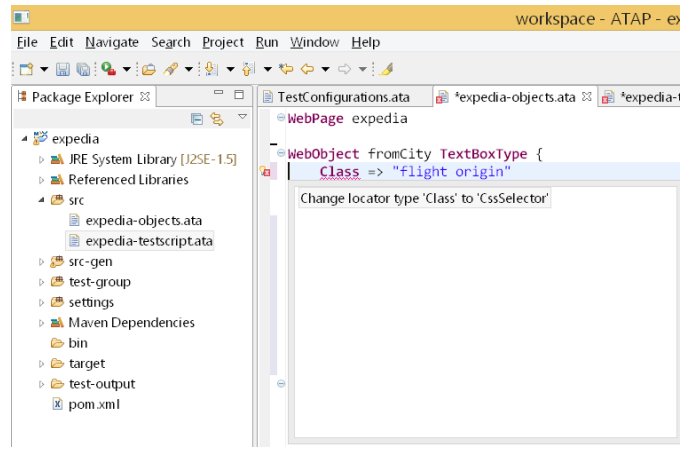

**Figure 9. Validation errors caught at creation time.**

*E. Code Generator*

The code generator module converts the test script written in the English-like syntax into Java code using Selenium WebDriver. The code generator is written in the Xtend [1] programming language. The generated code follows the Page Object Model design pattern [11], which is considered as a best practice in Selenium programming.

The code generator also abstracts the browser specific idiosyncrasies from the user. For example, the Java code which is able to successfully check the 'visibility' of a particular web element is different for Internet Explorer than for other browsers. For the tester authoring the automation script in ATAP, the English-like command is the same irrespective of the browser used. Numerous other cases as encountered by Selenium programmers have been incorporated into the code generator, thus helping new testers to create automation scripts faster.

The code generator is currently designed to generate Java code using the WebDriver API. Conceptually, the ATAP platform can be designed to work upon other automation platforms like HPE UFT or IBM RFT. This capability, when built, will allow a tester to abstract from the specific programming languages needed by the different tools. Thus, there wouldn't be a need to re-skill oneself in another tool.

*F. Execution of the test script*

Once the code generator creates the Java code from the DSL statements, the execution can be invoked directly from the test script editor – i.e. ATAP links the Java executable class with the corresponding English-like test script. This allows a tester to start the execution of the test script from the English-like script and need not navigate to the generated Java code. ATAP uses the test harness of TestNG [16] for the execution.

*G. Reporting and Linking errors*

A significant amount of effort in ATAP was spent on linking the Java code with the corresponding English-like statements. Through this, when there is an error in the Java code during runtime, the error is reported at the corresponding English-like statement. This allows the tester to stay only at the DSL level and does not require navigation to the Java code. Further, the linkage also allows the tester to debug from the DSL statements itself.

IV. CASE STUDY

We have deployed ATAP in a real-life software testing project. The project is in the hospitality industry and has various web sites around its different functions. The project has a set of automation engineers (who use Selenium WebDriver) and a set of manual testers. Both the automation engineers and the manual testers used ATAP and reported results.

In this section, we will report the quantitative measures of performance of ATAP. During the process of this exercise, we have also observed and received various qualitative feedback, which we report in the subsequent section.

The performance of ATAP was measured in terms of coverage (can ATAP automate the different scenarios in the project without needing Java code) and effort savings (the amount of time that is saved in using ATAP instead of Selenium WebDriver).

The project team selected a set of 8 test scenarios which spanned different levels of complexity. The scenarios required looping constructs, using regex, a database call and usage of variables. All the tests were conducted on the Internet Explorer browser.

The results were measured through three metrics. Every test scenario was authored in ATAP by a) The ATAP tool developers (who can be considered as tool experts); b) the automation engineers at the project (who can be considered experts of automation using Selenium, but are new to ATAP) and c) the manual test engineers at the project (who can be considered as new to test automation and new to ATAP). The manual test engineers had some knowledge of programming and also had some exposure to other automation tools (particularly RFT and using the capture & replay function).

For each test script, we measured the coverage and the effort savings achieved.

A. *Coverage*

In total, 8 different scenarios were automated. All the 8 scenarios were successfully automated using ATAP by the tool developers and by the project automation engineers showing 100% coverage. There was no need to use Java code for any of the scenarios.

The project's manual test engineers on average could automate 71% of a test scenario. This is an encouraging number as the manual testers attempted the authoring of automation scripts for the first time. The manual test engineers required help in using certain features of the tool – a) initial set-up of ATAP and ways to navigate Eclipse; b) help in identifying the grammar for cases needing loops; c) help in using specific grammar in validation. We present more observations on the usage of ATAP by the manual testers in the subsequent section. The total coverage achieved through the use of ATAP is shown in Table 1.

**Table 1. Amount of automation coverage achieved using ATAP.**

| Test Script Number | Test Script Complexity | Amount of coverage achieved by ATAP tool developers | Amount of coverage achieved by project's Automation engineers | Amount of coverage achieved by project's manual test engineers |
|---|---|---|---|---|
| 1 | High | 100% | 100% | Not attempted |
| 2 | High | 100% | 100% | 75% |
| 3 | High | 100% | 100% | 65% |
| 4 | High | 100% | 100% | 60% |
| 5 | Medium | 100% | 100% | 70% |
| 6 | Medium | 100% | 100% | 60% |
| 7 | Low | 100% | 100% | 70% |
| 8 | Low | 100% | 100% | 100% |
| Average | - | 100% | 100% | 71% |

B. *Effort*

The effort savings were calculated by accurately measuring the amount of time the ATAP developers and project engineers took to author the test scripts in ATAP. These metrics were compared against the estimate of the amount of time it takes the project automation engineer to author test scripts in Selenium. The detailed results are shown in Table 2. We did not use the time taken by the manual test engineer since the test scripts could not be completed entirely.

**Table 2. Effort savings through the use of ATAP (time is measured in hours).**

| Test Script Number | Test Script Complexity | Estimate of the time to author test script | Actual Time taken | | % savings for the automation engineer |
|---|---|---|---|---|---|
| | | | ATAP tool expert | Project automation engineer | |
| 1 | High | 12 | 3 | 10 | 16.67% |
| 2 | High | 11 | 4 | 9 | 18.18 % |
| 3 | High | 13 | 3 | 8 | 38.46 % |
| 4 | High | 15 | 3.5 | 10 | 33.33 % |
| 5 | Medium | 11 | 3 | 8 | 27.27% |
| 6 | Medium | 13 | 4 | 10 | 23.07 % |
| 7 | Low | 8 | 2 | 7 | 12.5 % |
| 8 | Low | 8 | 4 | 6 | 25 % |
| Total / Average | - | 91 | - | 68 | 25.27% |

The effort measurement showed that in all the cases, the automation engineers and the tool developers were able to create the scripts faster than the estimated time. On average, the automation engineer could save 25% of effort through the use of ATAP.

As can be seen in Table 2, the tool developers were able to automate the scripts much faster than the project's automation engineers. We believe this to be primarily due to the familiarity with the tool and the grammar. We explore this aspect further in Section V.

## V. OBSERVATIONS & FEEDBACK RECEIVED

In the process of the evaluation of ATAP in the project, there were various observations as well as explicit feedback. We report these lessons learnt below.

### A. Usage of the DSL

The DSL of ATAP was used by the automation engineers of the project without needing any help. These engineers also remarked that a significant amount of Java code is auto generated particularly noting that the page object model is followed. The DSL editor, being built into the Eclipse framework, made the automation engineers comfortable. Features supported in the Eclipse framework including linking different files, proposals, views, console output and reports were already known to the automation engineers.

The manual testers using the DSL could on average create 71% of a test scenario. The manual testers found the Eclipse framework to be unwelcoming as first. The general feedback was that, although the English-like script is aimed at making automation user friendly, the usage of Eclipse as the underlying platform is unhelpful (Eclipse was considered as a programmer's tool).The manual testers also requested for help on numerous occasions including ways to create a project (ATAP uses Maven to create projects), ways to rename a file, how to navigate between views and where the reports are generated.

The manual testers also found the DSL related to actions (starting with the 'When' keyword) self-explanatory. However, the manual testers found the verification statements (starting with the 'Then' keyword) difficult to interpret. After digging a little deeper into this feedback, we believe the reason is attributable to two factors. The first factor, we believe, is that our DSL starting with the 'Then' keyword had too many options requiring the tester to go through a long list of options. We have now made the DSL for the verification statements succinct where there are multiple levels with shorter options. The second factor, we believe, is quite subtle. We noticed that most of the manual test scenarios (which are written in natural language) tend to have fairly detailed action statements (such as - *select the dropdown value of 'Mr'*), however, the expected results were left at a high level (such as - *verify the room is booked*, or *verify the displayed results are correct*). Transforming this high level statement into specific commands required effort in creating a design for the test script (atleast as a mental model). For example, to verify the displayed results are correct, required the creation of variables and saving the relevant input data. Such verifications also needed some logical programming skills.

Overall, both the automation engineers and the manual testers appreciated the DSL where the English-like statements makes the creation of the automation script easy.

### B. Testing & Debugging of the automation script

During the course of the evaluation, we noticed that a large amount of time spent in the creation of the test script, actually went towards testing the script – i.e. checking that the script written is actually performing as expected. Further, when the test script failed at a particular step, the tester needed to re-execute the test script from the beginning after making the edits. This resulted in a lot of time being spent before a perfectly working test script was obtained. The need to test and debug was the core reason why the project automation engineers took almost twice the amount of time that the ATAP tool developers took (see Table 2).The ATAP tool developers knew the best command to use to automate a step, while the automation engineers attempted a few different methods before finding the one that works.

We believe, a significant advancement can be made if we can help reduce the amount of time it takes to test and debug a test script. There are a few ways in which this can be achieved. The ways in which a Selenium WebDriver code can fail are known apriori (for example, the exceptions include – `NoSuchElementException`, `ElementNotVisibleException`, etc.). In the generated code of ATAP, these exceptions can be caught and the program execution can be paused – i.e. the execution of the test script can be stopped at the occurrence of a set of runtime exceptions. This will allow a tester to edit the code (albeit the code can be edited in a constrained manner) and resume the execution. In this way, the execution of the test script need not be done from the beginning.

A second way to hasten the testing & debugging is to capture known cases of failure and suggest ways to overcome them. For example, when a `NoSuchElementException` is encountered, the existence of possible cases can be checked – like the presence of a frame, etc. This can then be prompted to the tester for correction.

### C. Standard tool features

ATAP was often compared with other commercial testing tools like Tricentis Tosca [18]. An oft mentioned drawback of ATAP was the lack of 'standard tool features' as available on other tools. The features expected were detailed reporting (ATAP uses the reporting framework of TestNG), validation of text in multiple languages, validating XML files and automatic capturing of web object locators.

It was felt that these features are quite rudimentary and are needed to make ATAP a complete tool. As tool innovators, we were on the other hand, more interested in developing useful DSL constructs and efficient auto-generated Java code. We believe a good strategy, going forward, would be to build the innovation around an existing robust test automation platform instead of building a tool from scratch. This will allow the project teams to reap the benefits from the innovation without having to sacrifice other basic features.

## VI. CONCLUSION

In this paper, we have presented a tool to accelerate the adoption of test automation – called ATAP. ATAP uses a domain specific language where testers, without programming skills, can create automation test scripts using a simple English-like syntax. The syntax follows the popular grammar of Gherkin. Internally, ATAP converts the statements written in the DSL to Java code using Selenium WebDriver APIs to drive the browser.

ATAP was deployed in an actual project and the results were measured. It was seen that the project's automation engineers were able to use ATAP effectively and it reduced their manual effort by 25%. The manual testers of the project were also able to use ATAP and they could successfully automate 71% of a test scenario.

There were numerous other observations and feedback received from the case study. It was observed that a significant amount of time was spent in actually testing and debugging the test script. We believe there is scope to considerably reduce this time through exception handling techniques and best practices.

The concept and the implementation of ATAP were very well received by the project team. Current efforts are in deploying ATAP in other scenarios within the project and improve the tool's capability particularly around debugging and testing of test scripts.